Alloyed cementite (Fe–Ni–Cr)$_3$C: structure and hyperfine field from DFT calculations and experimental comparison

L.V. Dobysheva

lyuka17@mail.ru    phone: +7(3412)728779
Institute of Physics and Technology, Udmurt Federal Research Center,
Ural Branch of Russian Academy of Sciences
34 Tatiana Baramzina str., Izhevsk, 426067 Russia

ABSTRACT
The alloying elements introduced into carbon steel to enhance its mechanical properties also diffuse into cementite (Fe$_3$C) particles, modifying their characteristics and thereby influencing the overall performance of the steel. This study employs density functional theory (DFT) calculations to investigate cementite doped with Ni and Cr which exhibit contrasting effects. The preferred lattice sites of impurity atoms were determined through a comparison of calculated and experimental structural parameters. The formation mechanism of the hyperfine magnetic field (HFF) and its correlation with atomic magnetic moments were systematically investigated. The validity of common approximations in Mössbauer spectroscopy analysis was evaluated for the cementite system. HFF distribution functions were modeled using calculated values and compared with experiments.

## 1. Introduction

In our previous study [1], we employed density functional theory (DFT) calculations to investigate the alloying behavior of Cr and Ni in cementite. These elements exhibit contrasting characteristics: Cr acts as a carbide-forming element while Ni does not, and they display antiferromagnetic and ferromagnetic ordering in cementite, respectively. Our DFT results [1] demonstrated that both Cr and Ni preferentially occupy the general position (site II) due to its lower energy configuration. We conducted detailed analyses of atomic magnetic moments (Fe, Ni, Cr) and the resulting net magnetization, which successfully explained the experimentally observed magnetization trends with varying impurity concentrations.
In a subsequent investigation [2], structural parameters previously obtained from mechanically alloyed and annealed (Fe-Ni-Cr)$_3$C samples were systematically analyzed.
In this study, we:
1) Determine the preferred impurity positions in mechanically alloyed samples through comparison of experimental lattice parameters from [2] with our current computational results;
2) Investigate the doping-dependent formation of hyperfine magnetic fields (HFF) at Fe nuclei using DFT calculations, examining their correlation with atomic magnetic moments;
3) Evaluate common approximations employed in Mössbauer spectral analysis and assess their validity for cementite systems;
4) Simulate the distribution function P(H) from calculated HFF values and compare with experimental P(H) from references [3-8].

## 2. Details of calculation

The method is described in [1]. Here, we outline only the basic principles and notations.
Doped cementite is modeled as an ordered crystal with infinitely repeating unit cells (UC). Since many local properties depend on the nearest atomic environment, this approach can also describe disordered systems.
The UC of pure cementite contains 16 atoms: 4 carbon atoms and 12 iron atoms in two nonequivalent positions — 4 in "special" (I) and 8 in "general" (II) sites. In undoped $Fe_3C$, a Fe I/II atom has two/three C atoms at 1.96–2.39 Å, 12/11 Fe neighbors at 2.45–2.68 Å, and 1/0 C atom at 2.82 Å. Beyond these distances, the next nearest atoms start at 3.48/3.58 Å, separated by a clear gap.

In alloyed cementite, one or two Fe atoms are replaced by Ni or Cr, giving an impurity concentration c = 8.3% or 16.7% in the metal sublattice. These values represent the fraction of Fe atoms substituted by dopants. The doped sites are labeled CrI, CrII, NiI, and NiII. Figure 1 compares two UCs of pure cementite (left) with two Ni substitution configurations (right).

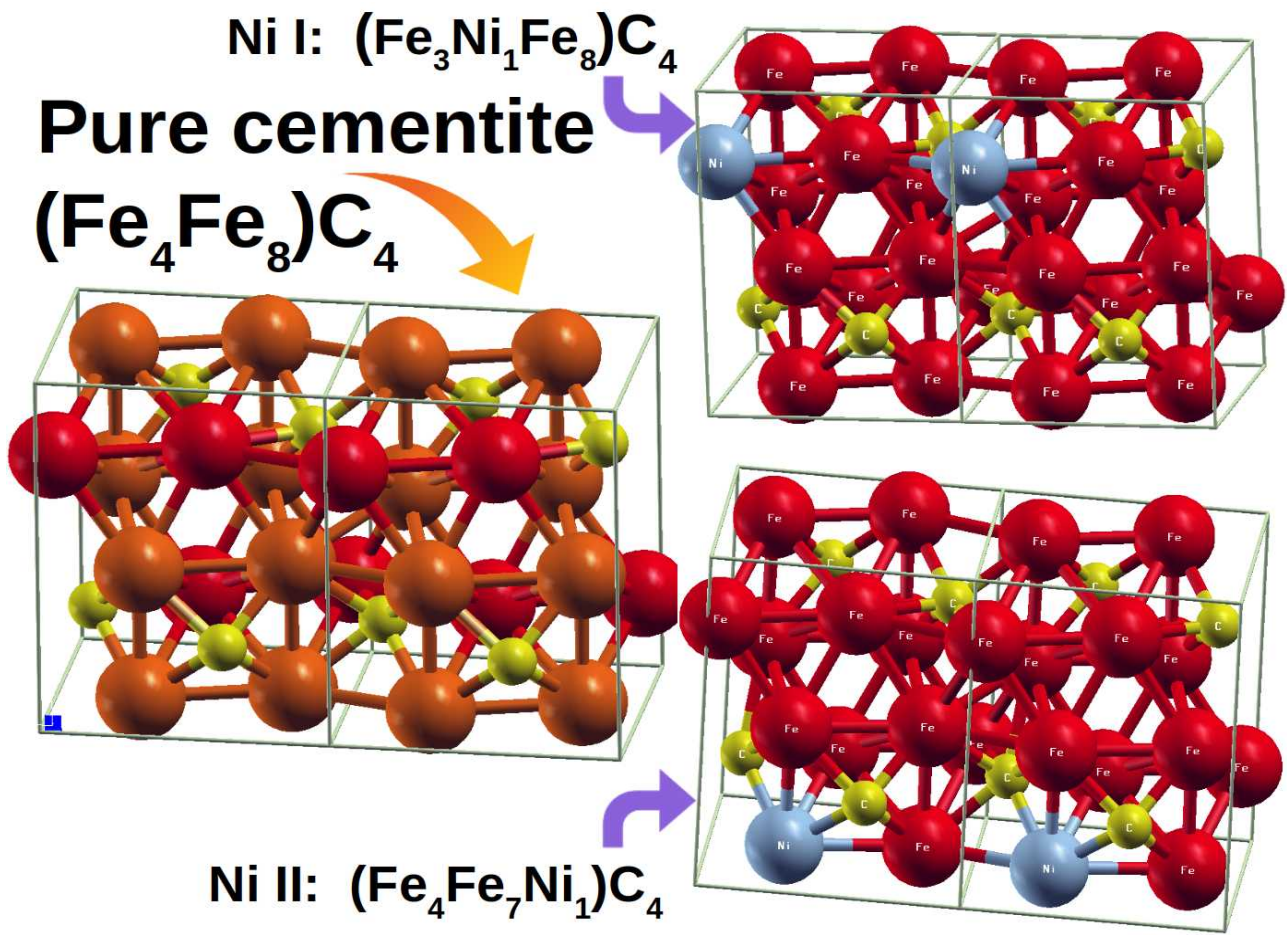

Fig. 1. Two UCs of pure $Fe_3C$ cementite (left) and two variants of Ni-doped cementite (right). The red/orange balls on the left represent Fe I/II atoms, the yellow/blue balls represent C/Ni atoms.

A 16.7% impurity concentration (two impurities per UC) creates many possible configurations. We focus on the most meaningful cases - where impurities are either close or far apart. For Cr-doped cementite ($Fe_{10}Cr_2C_4$), we examine four configurations: CrICrIfar (both Cr atoms are in sublattice I), CrICrIIclo, CrIICrIIfar, and CrIICrIIclo. Similarly, four configurations were studied for Ni-doped cementite ($Fe_{10}Ni_2C_4$). For the mixed case ($Fe_{10}CrNiC_4$, 8% of Cr and 8% of Ni), five configurations were considered since CrINiIIclo differs from NiICrIIclo.

Although we only study 18 configurations, they contain 186 non-equivalent atoms with distinct nearest-neighbor environments (NN).

Doping distorts the symmetry, making the classification of atomic positions into general or special less strict. However, since atomic environments remain similar to pure cementite (see below), we still assign metal atoms to sublattices I or II.

The above mentioned distance gap persists in doped cementite, allowing clear definition of NN. We consider 15/14 atoms as NN for initial I/II positions.

Cementite's complex structure is particularly interesting for analysis due to its many non-equivalent atomic sites with distinct NN configurations.

The DFT calculations of the electron density were performed using the full-potential linearized augmented plane wave (FP LAPW) method implemented in the WIEN2k software package [9]. The exchange and correlation potential was calculated in the Perdew-Burke-Ernzerhof parameterization of the generalized gradient approximation (GGA) [10]. The calculation parameters were chosen using the standard approach with a convergence check on the parameters: muffin-tin (MT) sphere radii $R_{MT}$ = 2.04/2.03/2.04/1.59 a.u. for Fe/Ni/Cr/C atoms, respectively. The expansion of the wave functions of valence electrons in atomic spheres was limited by the angular momentum $l_{max}$ = 10. The expansion of wave functions outside the spheres was limited by the cutoff vector $K_{max}$ = 7.0/$R_{MT}$(C); the charge density was expanded in a Fourier series to a vector with a maximum value $G_{max}$ = 12 $Ry^{1/2}$; a grid of 1000 k-points was used in the Brillouin zone.
In our calculations (consistent with standard DFT approaches), electron levels are separated into core and valence states - an important distinction for hyperfine parameter calculations. This separation accounts for their different spatial localization: core electrons are confined within atomic spheres (1% charge accuracy) and valence electrons extend beyond MT spheres. We used these core configurations: $1s^22s^22p^63s^2$ for Fe and Ni, $1s^22s^22p^6$ for Cr, and $1s^2$ for C. The WIEN2k method treats core electrons fully relativistically and valence electrons scalar relativistically. The equilibrium lattice parameters and atomic positions were first determined.
The atomic magnetic moments were calculated by integrating the spin density ($\rho\uparrow - \rho\downarrow$) over the MT sphere. The hyperfine magnetic field was calculated as HFF=$H_{val}$+$H_{core}$ considering only valence and core electron contributions. They are proportional to the spin density ($\rho\uparrow - \rho\downarrow$) of the corresponding electrons, integrated over the Thomson sphere (slightly larger than the Fe nucleus). The isomer shift (IS) is derived from electron density ($\rho\uparrow + \rho\downarrow$) at the Fe nucleus center, relative to a reference: $\rho - \rho_{etalon}$.
For validation, calculations for pure cementite [11,12] and (Fe-Ni)$_3$C [13] were reproduced using the same computational methodology as employed in this work.

**3. Results and discussion**

**3.1 UC volume and lattice parameters**

In [1], we demonstrated that for both Ni and Cr, sublattice II is energetically more favorable. This work aims to identify traces of preferential impurity arrangement in sublattice II using experimental data from [2] (samples prepared by mechanical alloying (MA) followed by annealing). In [2], the experimental lattice parameters were compared with reference cementite parameters from [14], where Fe$_3$C was extracted from Fe-C ingots. The study showed that even annealing unalloyed mechanically synthesized cementite at different temperatures does not yield the standard Fe$_3$C parameters {a, b, c} = {4.523, 5.089, 6.743} Å [14], possibly due to insufficient annealing time. Since this work focuses on the influence of impurities, all calculated values are referenced against calculated unalloyed cementite {a, b, c} = {4.494, 5.042, 6.722} Å, while experimental data are compared to experimental unalloyed mechanically synthesized cementite from [2] {a, b, c} = {4.526, 5.089, 6.751} Å.
Figures 2–4 present the experimental changes in a, b, c, and V from [2] alongside our calculations for Cr-, Ni-, and Cr-Ni-alloyed cementite.

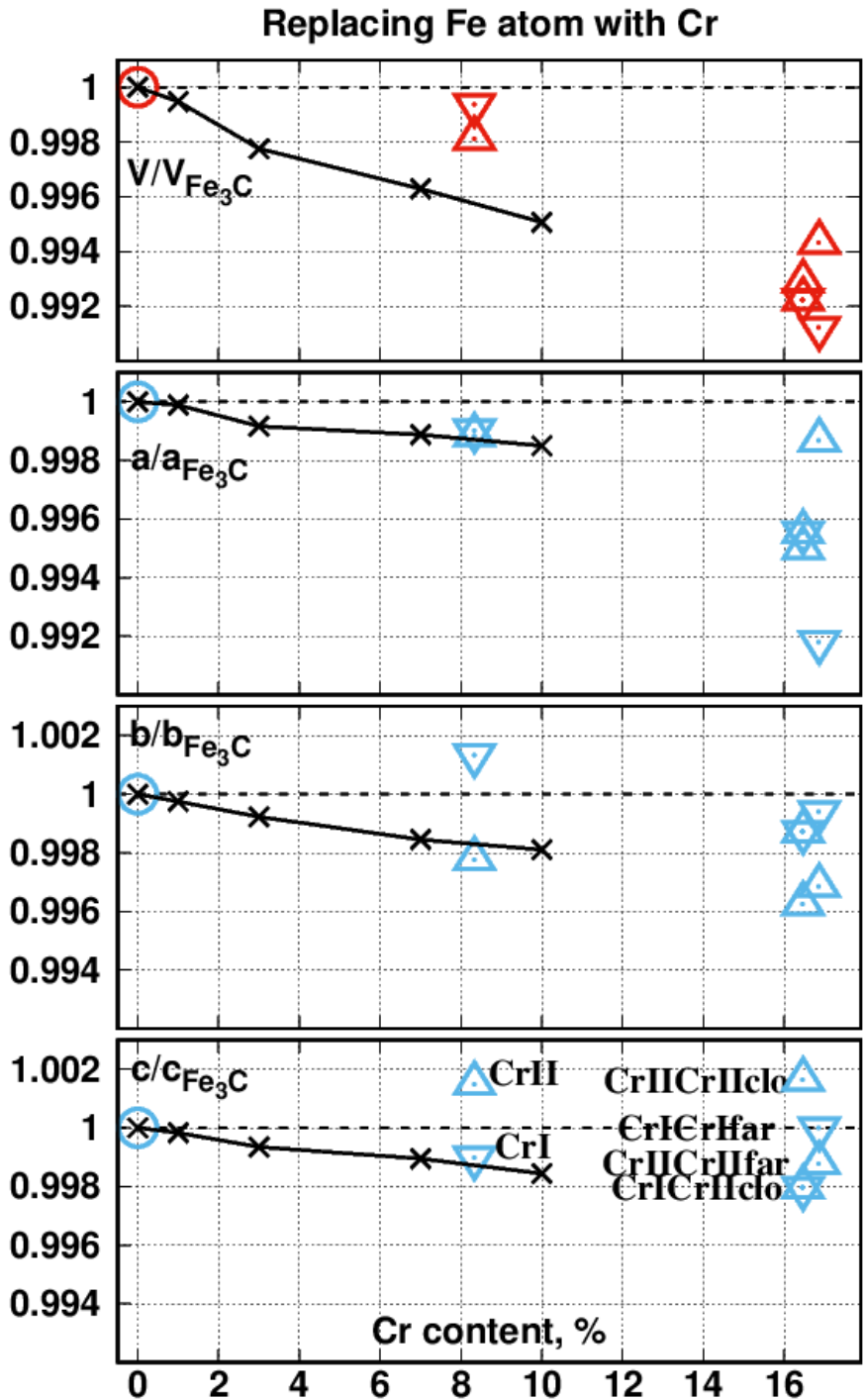

Fig. 2. Variation of a, b, c, and V with Cr content in (Fe-Cr)$_3$C. Calculated data are denoted as down/up triangles for Cr in sublattice I/II and hexagrams for Cr in both sublattices; experimental data from [2] are denoted as crosses with a line as a guide for the eye. Note: Points at 16.67 at.% are shifted slightly horizontally for clarity.

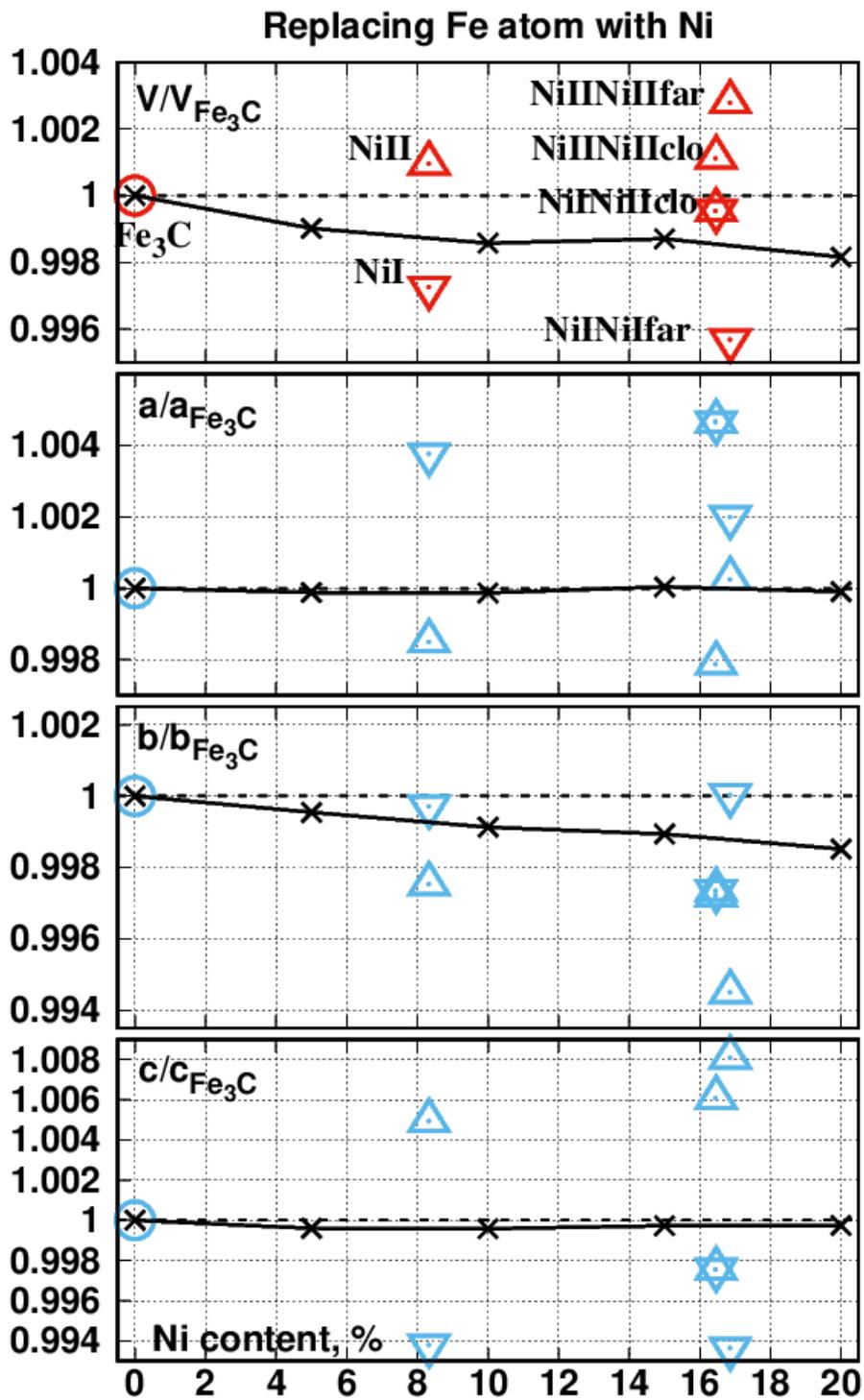

Fig. 3. Variation of calculated and experimental [2] a, b, c, and V with Ni content in (Fe-Ni)$_3$C (notations as in Fig. 2). **Note:** While calculated data for Ni show clear separation in sublattices I and II for a, c, and V, experimental values are close to averages, showing no site preference.

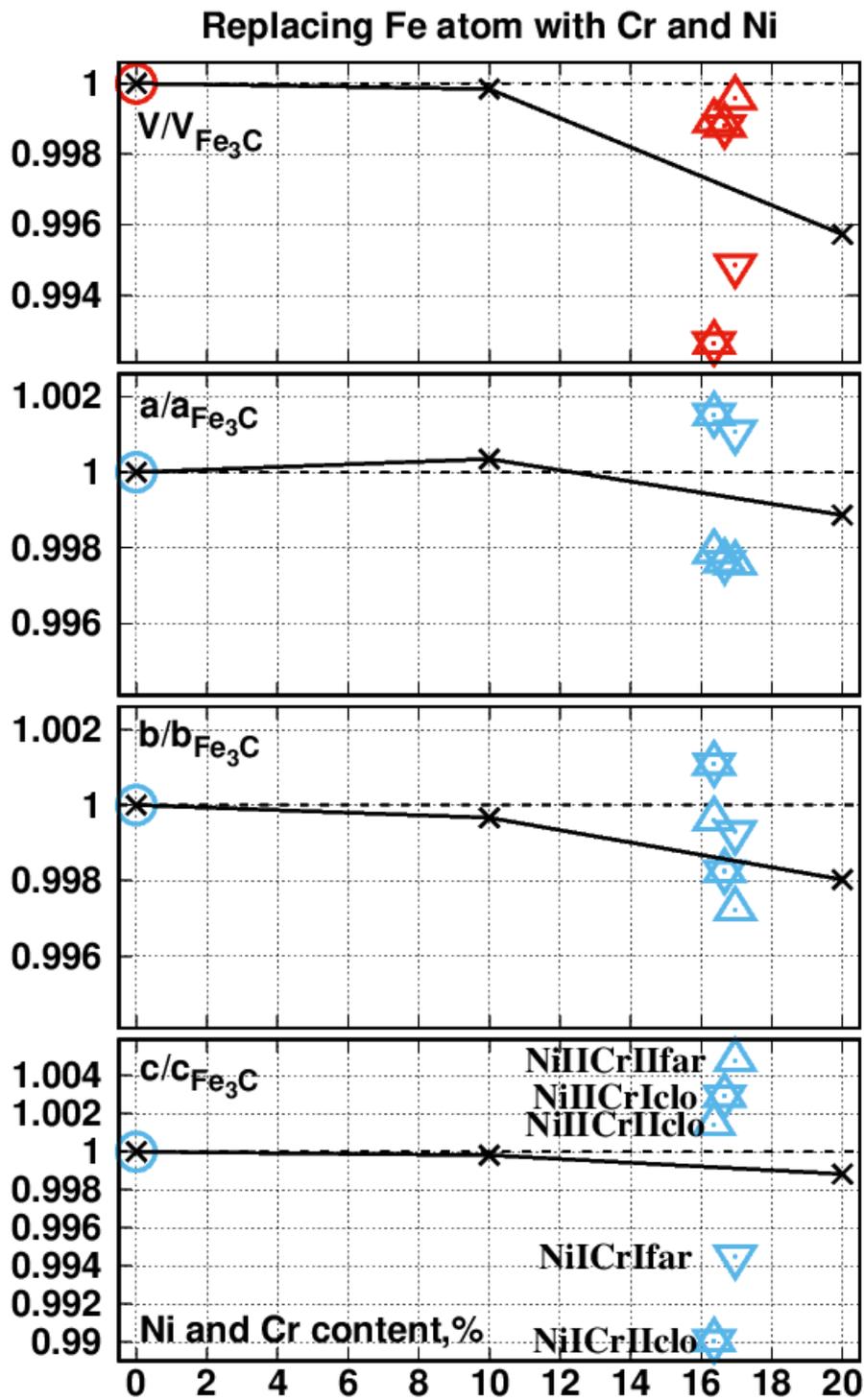

Fig. 4. Variation of calculated and experimental [2] a, b, c, and V with total Ni+Cr content in (Fe-Ni-Cr)₃C (notations as in Fig. 2). Both calculated and experimental data are shown for equal Cr and Ni concentrations. **Note:** As in Fig. 3, while a, c, and V clearly distinguish Ni I and II positions in calculations, experiments show no Ni site preference.

Fig. 2 reveals that both experimental and calculated UC volume reduction under Cr doping is due to changes in all three lattice parameters. Fig. 3 demonstrates that the five times smaller volume decrease with Ni doping arises solely from parameter b variation, with excellent theory-experiment agreement. Fig. 4 shows that Ni-Cr co-doping effects approximately combine the individual Ni and Cr behaviors, with a fairly good agreement between calculation and experiment. While our limited set of ordered structures prevents rigorous statistical analysis, the calculations correctly capture the experimental trends in volume and lattice parameter changes.

Calculations in [1] revealed distinct effects of Ni impurities on UC volume: NiII increases volume and NiI decreases volume. This contrast is also evident in lattice parameters (Ni and NiCr systems, Figs. 3-4). However, experimental data (crosses) fall between sets I and II, showing no Ni site preference.
Both CrI and CrII impurities reduce UC volume [1]. Our results show all three lattice parameters decrease with CrI and CrII doping. Unlike the Ni system, the Cr data in Figs. 2 and 4 have smaller value spread, less separation between CrI/CrII cases, and limited experimental range (≤10 at.% Cr). These factors prevent conclusive determination of Cr's preferential occupation.
Conclusion: The high-energy MA method yields a random distribution of Ni across sublattices, and subsequent heat treatment fails to produce impurity redistribution. The Cr system data do not permit definitive determination of Cr atoms' preferential positions.

### 3.2 Hyperfine magnetic field

We focus on the hyperfine magnetic field and its relationship with atomic magnetic moment, nearest neighbor environment, number of impurities (N) in NN, and isomer shift.
Figure 5 presents the NN configuration, M, and HFF for Fe in $Fe_3C$, $Fe_{11}CrIC_4$ (Cr in sublattice I), $Fe_{11}CrIIC_4$ (Cr in sublattice II), $Fe_{11}NiIC_4$, and $Fe_{11}NiIIC_4$. The points on the x-axis in the upper part show the atoms located in sublattices I/II (4/8 metal atoms) of the corresponding compound. Their environment (the above-mentioned 15/14 atoms) appears vertically, arranged by distance. The lower part shows the corresponding Fe M and HFF values.

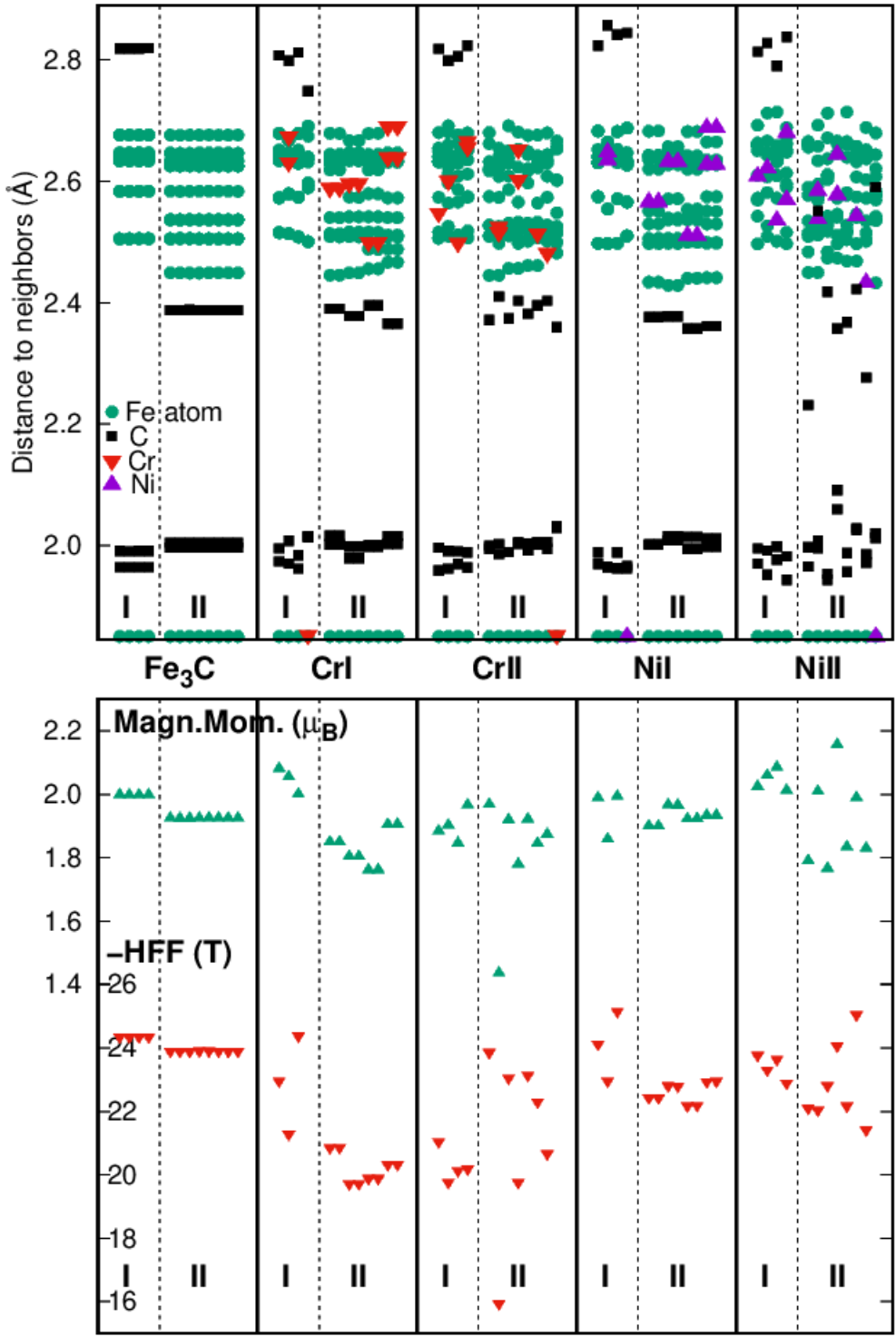

Fig. 5. Environment, magnetic moment, and hyperfine field of Fe in $Fe_3C$, $Fe_{11}CrIC_4$ (Cr in the I sublattice), $Fe_{11}CrIIC_4$ (Cr in the II sublattice), $Fe_{11}NiIC_4$, and $Fe_{11}NiIIC_4$.

We observe a general but non-strict correlation between M and HFF (see, for example, M and HFF of three Fe atoms in the I sublattice of CrI), though significant M deviations are reflected in HFF (like the second atom in CrII's sublattice II). The Supporting Information presents similarly structured data for $Fe_{10}Cr_2C_4$, $Fe_{10}Ni_2C_4$, and $Fe_{10}CrNiC_4$.

Structural and magnetic moment data can be also found in [1]. As demonstrated in Figure 5 and Figs. 1–3 in Supplementary Material, HFF variations due to the nearest environment do not strictly correlate with changes in M. We discuss this in more detail below.

### 3.3 HFF and atomic magnetic moment

Figure 6 displays the dependence of the HFF and its valence and core contributions on the magnetic moment.

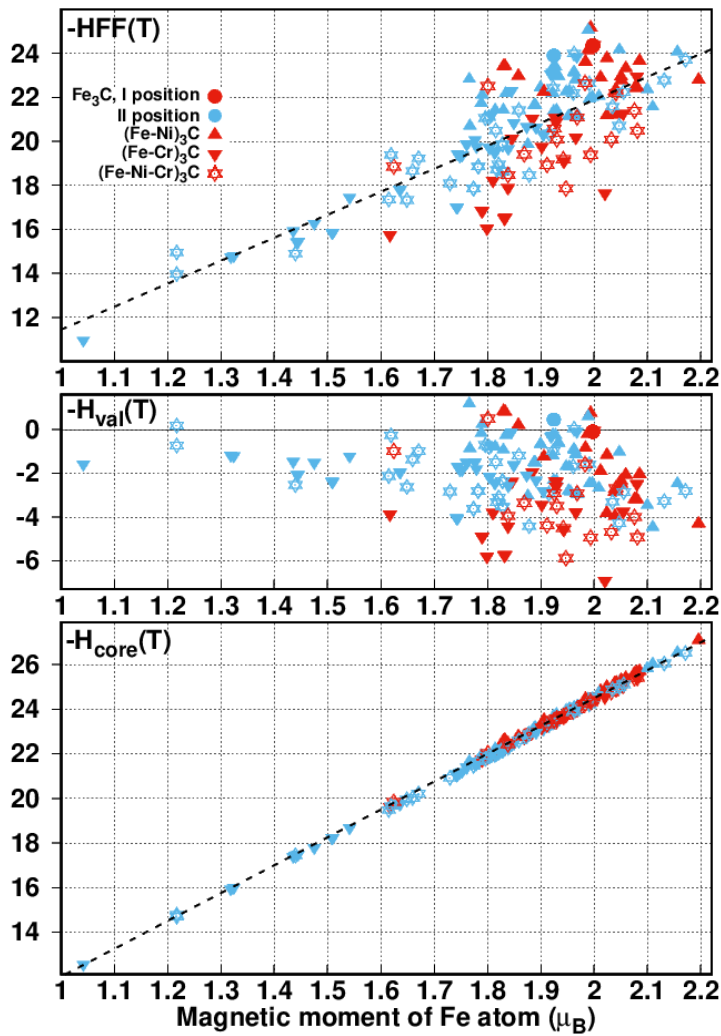

Fig. 6. HFF with valence and core contributions versus atomic magnetic moment.

As demonstrated in [15-17], the core contribution to HFF shows a strictly linear dependence on M, consistent with our results for alloyed cementite (Fig. 6, lower panel). This linearity indicates that the total spin density integrated over the MT sphere (equal to M) scales proportionally with the core spin density integrated over the much smaller Thomson sphere ($H_{core}$). This expected behavior arises because core electrons are localized within the MT sphere, practically do not interact with nearest atoms, and depend on environment only indirectly, namely through M.
The valence contribution (Fig. 6, center) shows both positive and negative values, depending on distance to the impurities and their magnetic moments - analogous to the long-range RKKY interaction [17]. Figure 6 reveals no consistent correlation between this contribution and M.
The total HFF ($H_{val} + H_{core}$) shows ±5 T variation at fixed M values, entirely due to $H_{val}$ dispersion. When analyzing Ni and Cr systems separately, the variation reduces to ±3 T - still too large for reliable discrete modeling of Mössbauer spectra.
For lower moments (<1.6 $\mu_B$), the dispersion decreases, because all these cases involve Fe II atoms with nearly identical environments: two Cr atoms occupy opposite NN positions around the Fe atom [1], resulting in equal valence contributions that maintain the $H_{core}$-M proportionality.
Conclusion: The hyperfine field and Fe atomic magnetic moment lack strict linear correlation due to valence electron contributions.

### 3.4 HFF and number of impurities in NN

Mössbauer spectra are often analyzed discretely, assuming the HFF depends on the number of impurities (N) in NN. Figure 7 shows the HFF as a function of N.

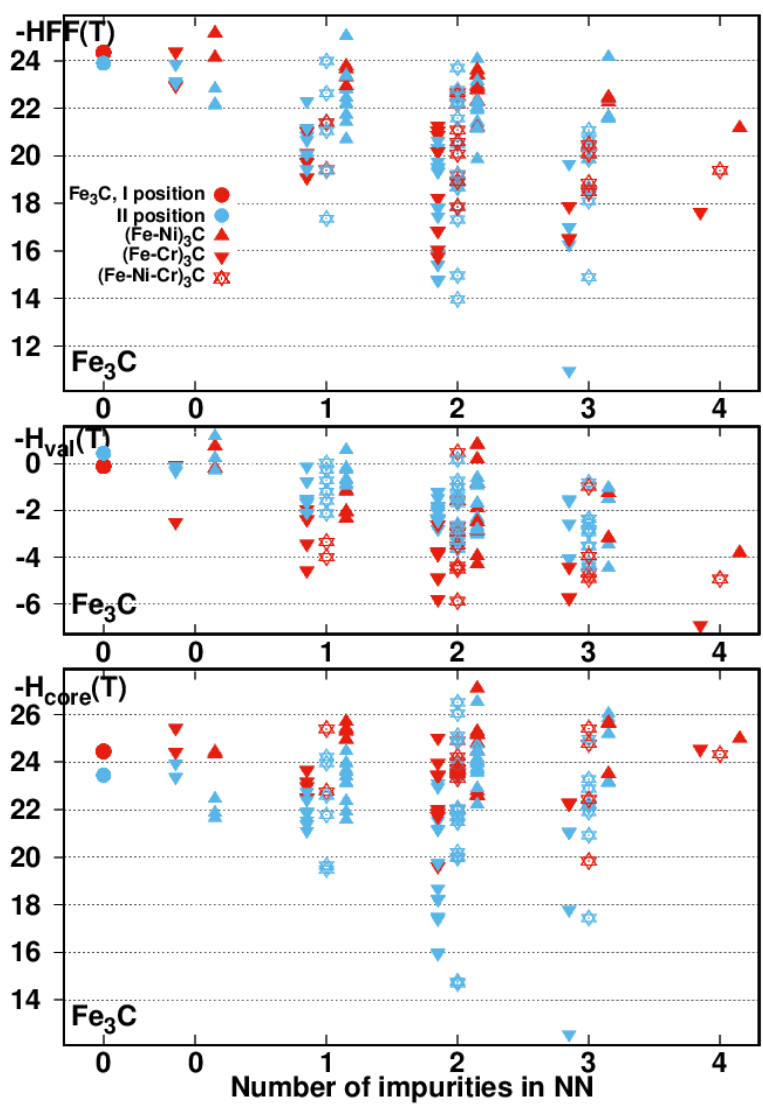

Fig. 7. HFF with valence and core contributions versus number of impurities in NN.

In (Fe-Ni)$_3$C, the average magnetic moment of Fe with N impurities in NN ($M^N$) remains nearly constant with N [1], but the average $H^N$ decreases with N (from 24 T to 21–22 T, Fig. 7, up triangles). This reduction comes solely from the valence electrons, as the average core contribution $H^N_{core}$ stays constant (since it scales with $M^N$, which does not change). In $(Fe-Cr)_3C$, the average $H^N$ drops more sharply (from 24 T to 17–18 T, down triangles), driven by both core and valence contributions. For $(Fe-Ni-Cr)_3C$, the average $H^N$ lies between the Ni and Cr cases (from 24 T to ~19 T, hexagrams).

Fig. 7 shows the case of four impurities in NN, providing a simple example of the M–HFF relationship: M is nearly identical across all three systems: $(Fe-Ni)_3C$, $(Fe-Cr)_3C$ and $(Fe-Ni-Cr)_3C$ [1]; $H_{core}$ also remains the same, as it is proportional to M; however, the total HFF differs significantly due to variations in the valence contribution (Table 1).

Table 1. Magnetic moment and HFF= $H_{core}$ + $H_{val}$ of Fe atom with 4 impurities in NN in the compounds CrIICrIIclo, NiIICrIIclo, and NiIINiIIclo

|             | M    | HFF   | $H_{core}$ | $H_{val}$ |
|-------------|------|-------|------------|-----------|
| CrIICrIIclo | 2.02 | -17.6 | -24.6      | 6.9       |
| NiIICrIIclo | 1.99 | -19.4 | -24.3      | 4.9       |
| NiIINiIIclo | 2.02 | -21.2 | -25.0      | 3.8       |

Like the magnetic moments [1], the HFF exhibits significant fluctuations around mean values due to variations in the Fe atom's local environment. Therefore, when analyzing Mössbauer spectra with a discrete model, mean HFF values must be used with sufficiently broad linewidths to account for this dispersion. The substantial HFF spread prevents the spectra of doped cementite from being treated as a simple superposition of NN-dependent subspectra.
At a low Cr percentage (8%), assuming a random distribution, the contribution of $H^0$ is ~38% with a fairly broad average of $<H^0>$= 23.5T ±1T, the contribution of $H^1$ is ~38% with $<H^1>$= 20.5T ±2T, and remaining contributions of other NNs with a maximum of $<H^2>$= 18T ±3T. For Ni-doped cementite, the $H^0$, $H^1$ and $H^2$ contributions cannot be resolved due to overlapping fields: $<H^0>$=23.5T ±0.7T, $<H^1>$=22.5T ±2T, $<H^2>$=22T ±2T.

Conclusion: In the studied alloyed cementite, the HFF exhibits significant variations even for identical numbers of impurities in NN. Consequently, Mössbauer spectra analysis using a discrete model requires a non-standard approach, where hyperfine parameter variations can only be properly described by employing broad spectral lines.

### 3.5 HFF and isomer shift

When processing Mössbauer spectra in continuous representation, the inverse problem is often simplified by assuming a linear HFF-IS dependence. Figure 8 shows the calculated HFF-IS correlation, including the corresponding contributions.

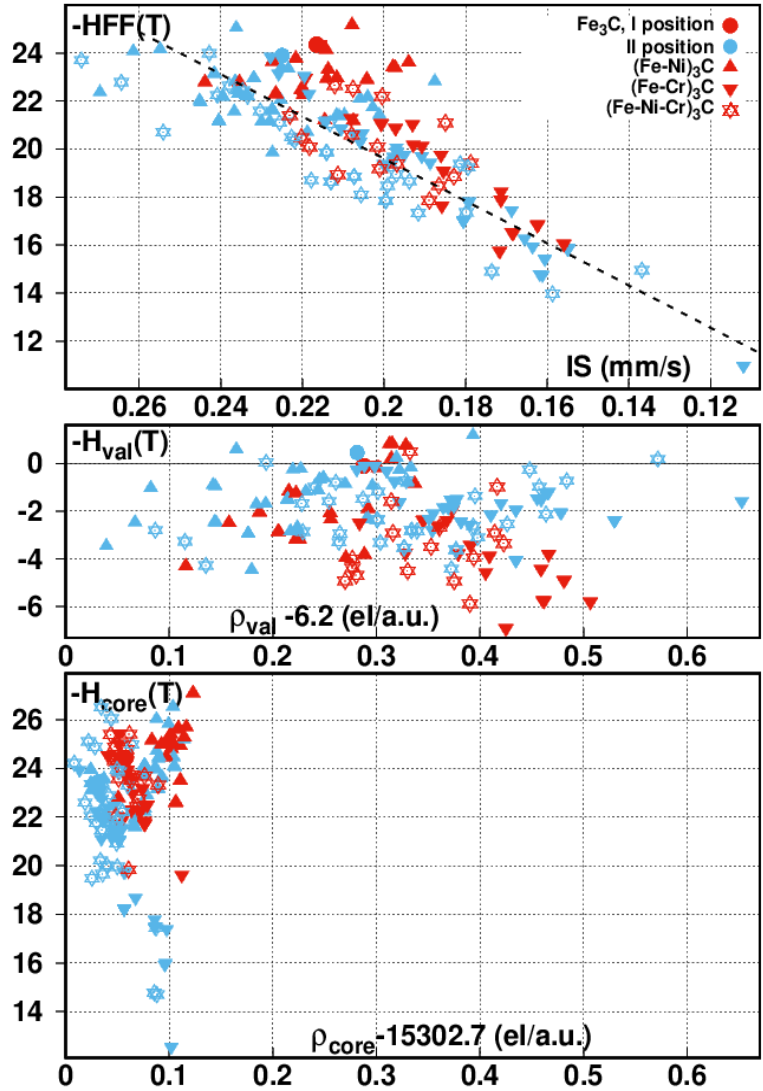

Fig. 8. HFF versus IS at Fe. Top: total values; middle: valence contributions; bottom: core contributions. Note: x-axis vary between panels.

For certain HFF values, the calculated IS variations (0.3-0.8 mm/s) significantly exceed the experimental measurement error (0.01 mm/s). The core IS contribution shows minimal dispersion (0.02 mm/s), as expected, since core electrons remain almost entirely within the MT sphere, making $\rho_{core}$ at its center nearly environment-independent; $IS_{core}$ is also largely insensitive to M (since IS ~ $\rho\uparrow + \rho\downarrow$), unlike $H_{core}$, which depends on spin density and shows substantial spread (Fig. 6).
The valence IS contribution (proportional to $\rho_{val}$ at the nucleus center) depends on the environment through interactions with neighboring atoms, as valence wave functions have long-range tails. $IS_{val}$ shows no correlation with $H_{val}$ (which depends on the spin density difference $\rho\uparrow - \rho\downarrow$ integrated over the Thomson sphere).

Conclusion: The HFF-IS relationship cannot be simply or unambiguously described. Using a linear HFF-IS approximation in Mössbauer spectra analysis may distort the results unpredictably.

## 3.6. Comparison of the calculated HFF with experiment

HFFs were calculated for 18 doped cementite systems (Fig. 5 and Figs. 1-3 in Suppl. Mater.), considering numerous non-equivalent Fe atoms in various NN. These results are compared with experimental data below. First, we correlate the HFFs with magnetization - a directly measurable property (see the experimental and calculated magnetization comparison in Fig. 5 of [1]). Fig. 9 presents the HFFs for all structures versus the total UC magnetization; magnetization trends versus impurity content can be also seen. Note that total magnetization includes, in addition to the magnetic moments of Fe atoms, those of Ni and Cr ($M_{Cr}=-0.3$ to $-1.0$ $\mu_B$ and $M_{Ni}=0.2-0.5$ $\mu_B$), which replace $M_{Fe} = 1.9-2.0$ $\mu_B$ [1].

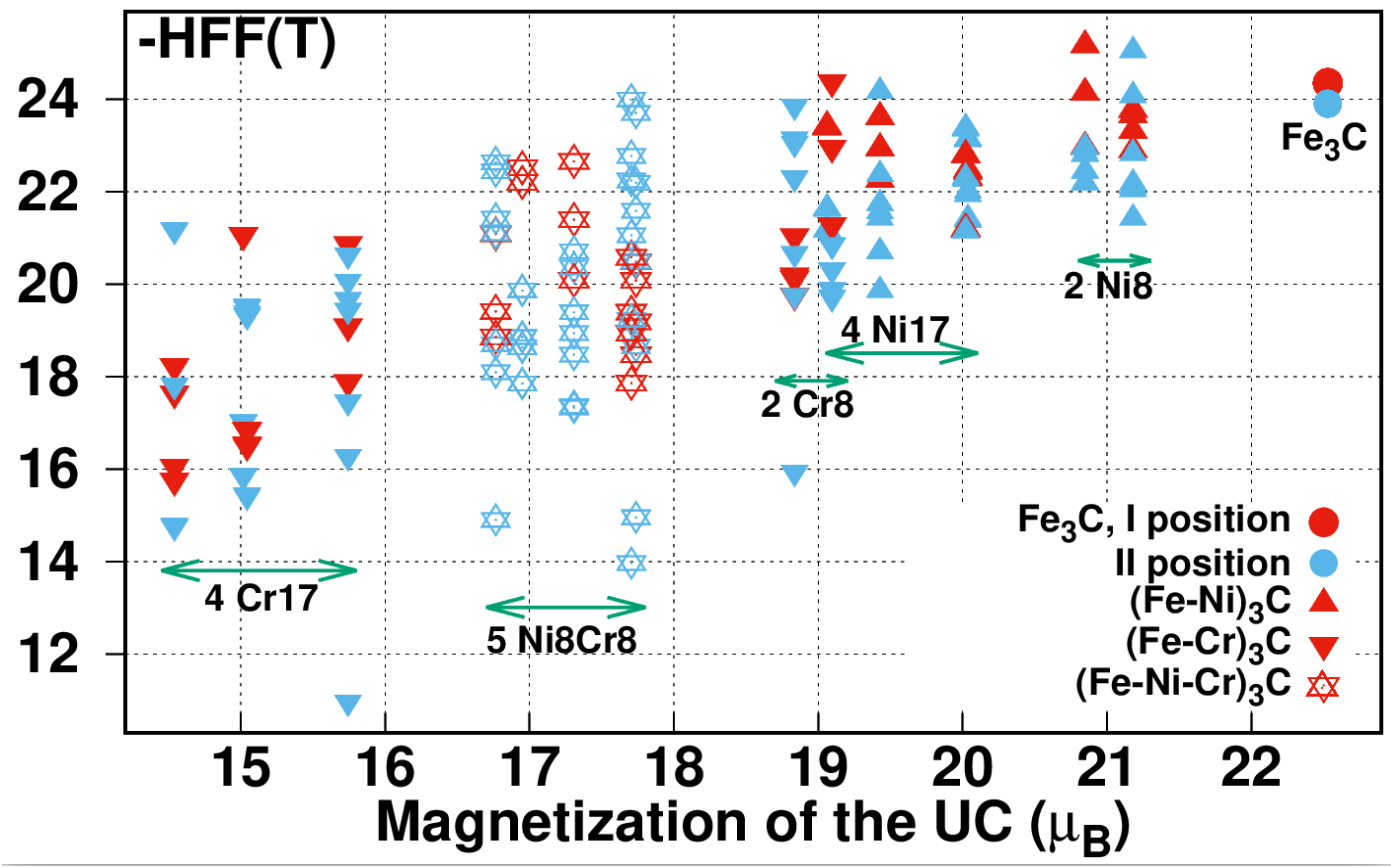

Fig. 9. HFF versus UC magnetization for 18 alloyed cementite configurations. Labels indicate number of configurations and their composition (e.g., '4 Cr17' = four configurations with 17% Cr; '2 Ni8' = two configurations with 8% Ni).

The numerous and diverse HFFs in cementite (Fig. 9) make individual spectral components unresolvable experimentally. Even for ordered alloyed cementite, a discrete model would fail to describe the spectrum adequately - without even considering the IS variations discussed in Section 3.5.

These calculations represent electronic ground states (T = 0 K), while experimental results on magnetization and HFF are affected by cementite's relatively low Curie temperature ($T_c$), which is doping-dependent. This effect was clearly demonstrated in [13] through Mössbauer spectra at 77 K and 300 K (see also Fig. 11 below). Cr doping reduces $T_c$ (decreasing with higher Cr content) [3-4]. Ni doping slightly increases $T_c$ compared to pure cementite [5].

To compare with Mössbauer experiments, we simulated HFF distribution functions P(H) for disordered systems with 8% or 17% impurities (simulations of Mössbauer spectra turned out to be less informative [13]). Each calculated Fe HFF was broadened into a Gaussian, with width matching experimental P(H) for $Fe_3C$ [3-4]. Contributions were weighted according to random impurity distributions using binomial function.
Fig. 10 compares calculated P(H) for Cr-alloyed cementite with experimental results [3-4]. Calculated curves are designated, for example, as "Cr17 calc", which was obtained from four UC configurations of Cr17 (Fig. 1 in Suppl. Mater.) with 40 HFFs shown in Fig. 9.

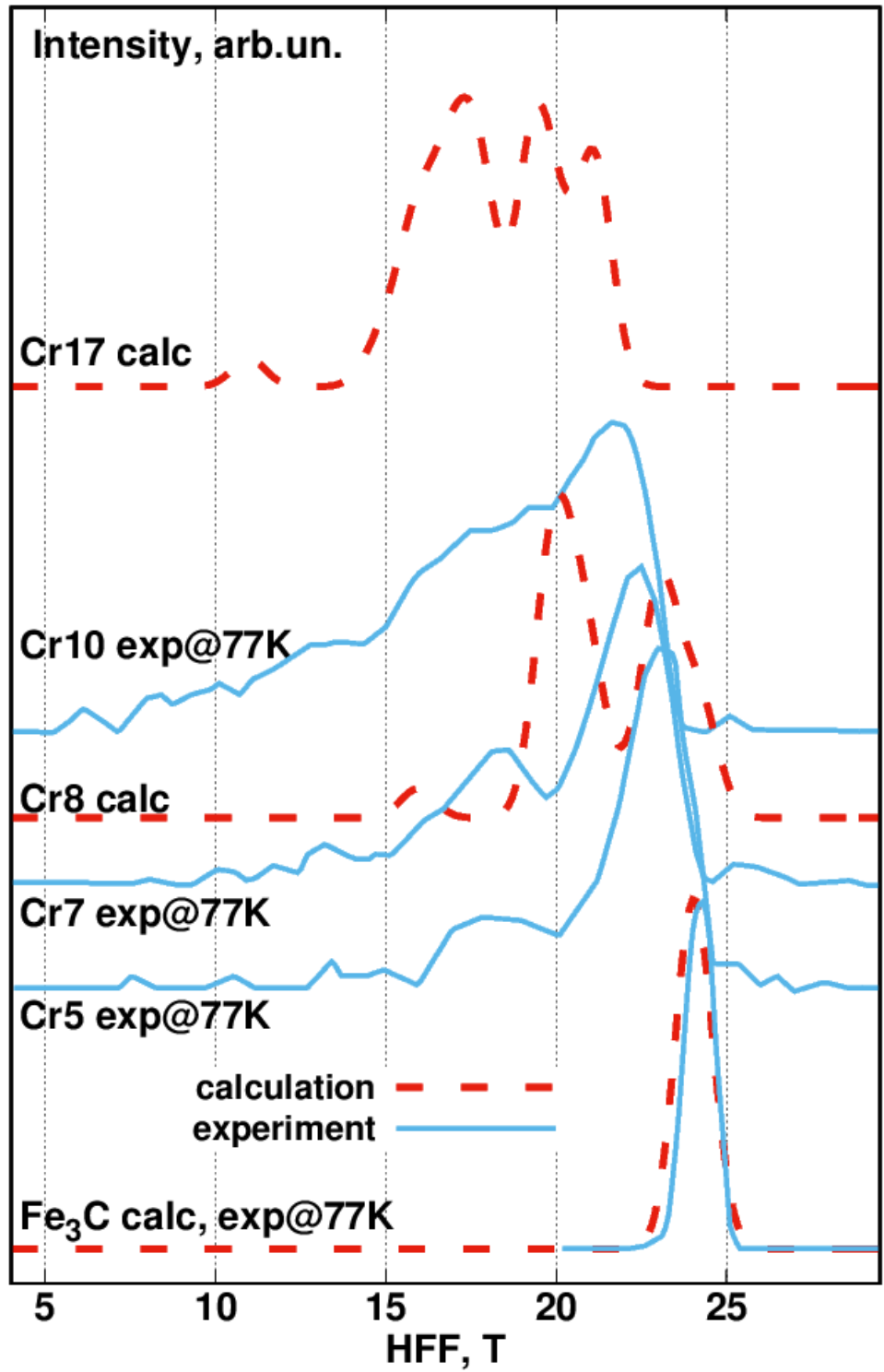

Fig. 10. Experimental [3-4] and simulated P(H) for Cr-doped cementite compared with pure $Fe_3C$. Curves are vertically shifted according to the Cr content (0, 5, 7, 8, 10 and 17%). Mössbauer experiments were carried out at 77 K. **Note:** Increasing Cr content

broadens P(H) and shifts it to lower HFF values.

As demonstrated in Section 3.5, the assumed linear HFF-IS relationship is invalid, potentially distorting the P(H) shape in inverse-problem solutions of Mössbauer spectra. Nevertheless, both experimental and calculated results show consistent trends: P(H) broadens and shifts to lower HFF with increasing Cr content; at lower concentrations, experimental (Cr5, Cr7) and calculated (Cr8) P(H) agree well. However the experimental Cr10 curve is wider than even the calculated Cr17 (discussed below). Fig. 11 shows the simulated P(H) for Ni-alloyed cementite and the data obtained in experiments [5].

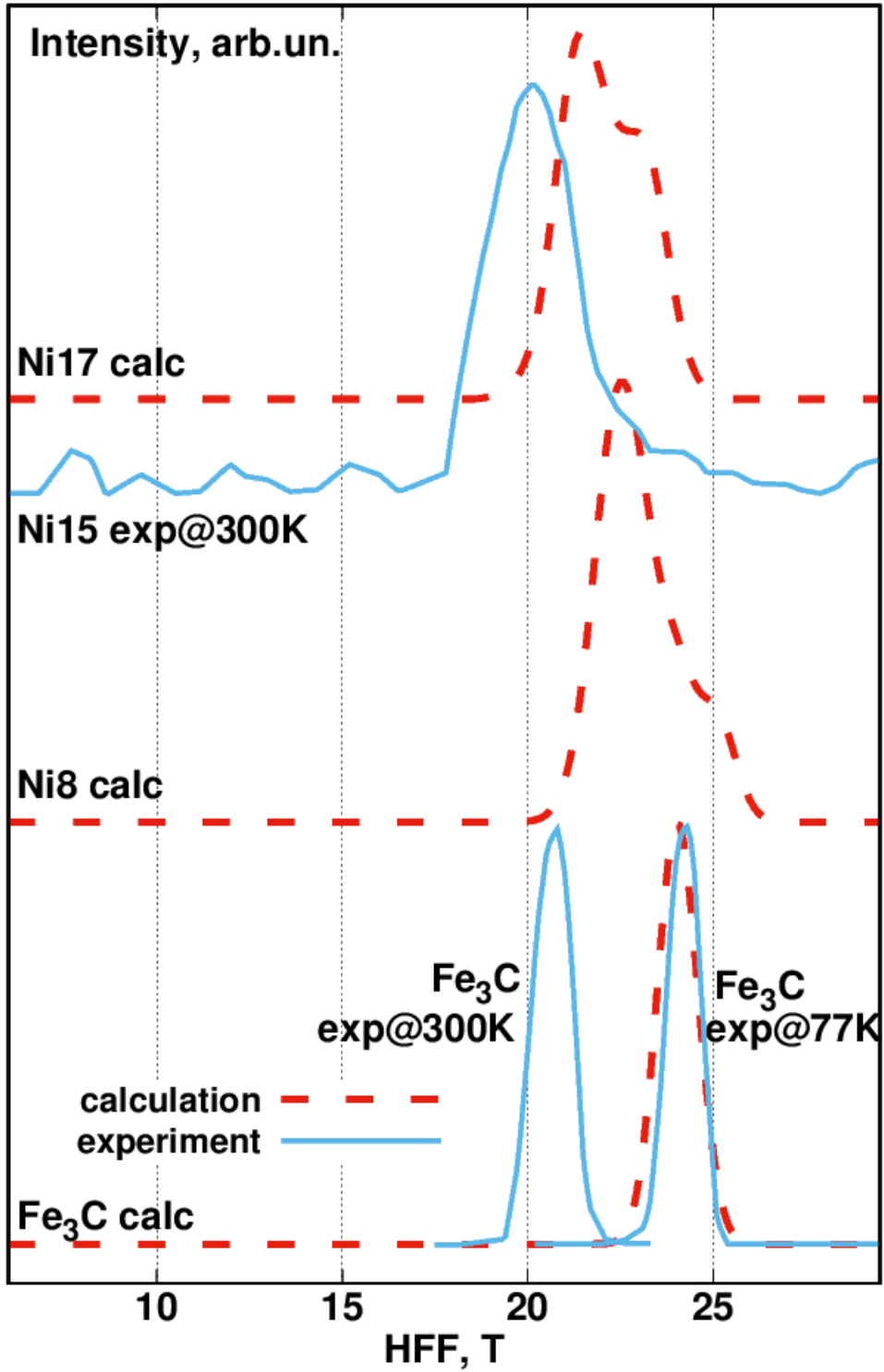

Fig. 11. Experimental [3,5] and simulated P(H) for Ni-doped cementite compared with pure Fe₃C (notations as in Fig. 10). **Note:** Increasing Ni content broadens P(H) and shifts it to lower HFF values.

The experimental P(H) of Ni15 is noticeably shifted compared to the calculated Ni17, due to the spectrum being recorded at room temperature. A similar shift is observed for pure cementite at 300 K vs. 77 K (Fig. 11). Accounting for this, both experimental and calculated P(H) shift toward lower HFF and broaden with increasing Ni content, though less than in the Cr system (Fig. 10).

Figure 12 compares the calculated P(H) for Ni- and Cr-alloyed cementite with experimental data [6-8].

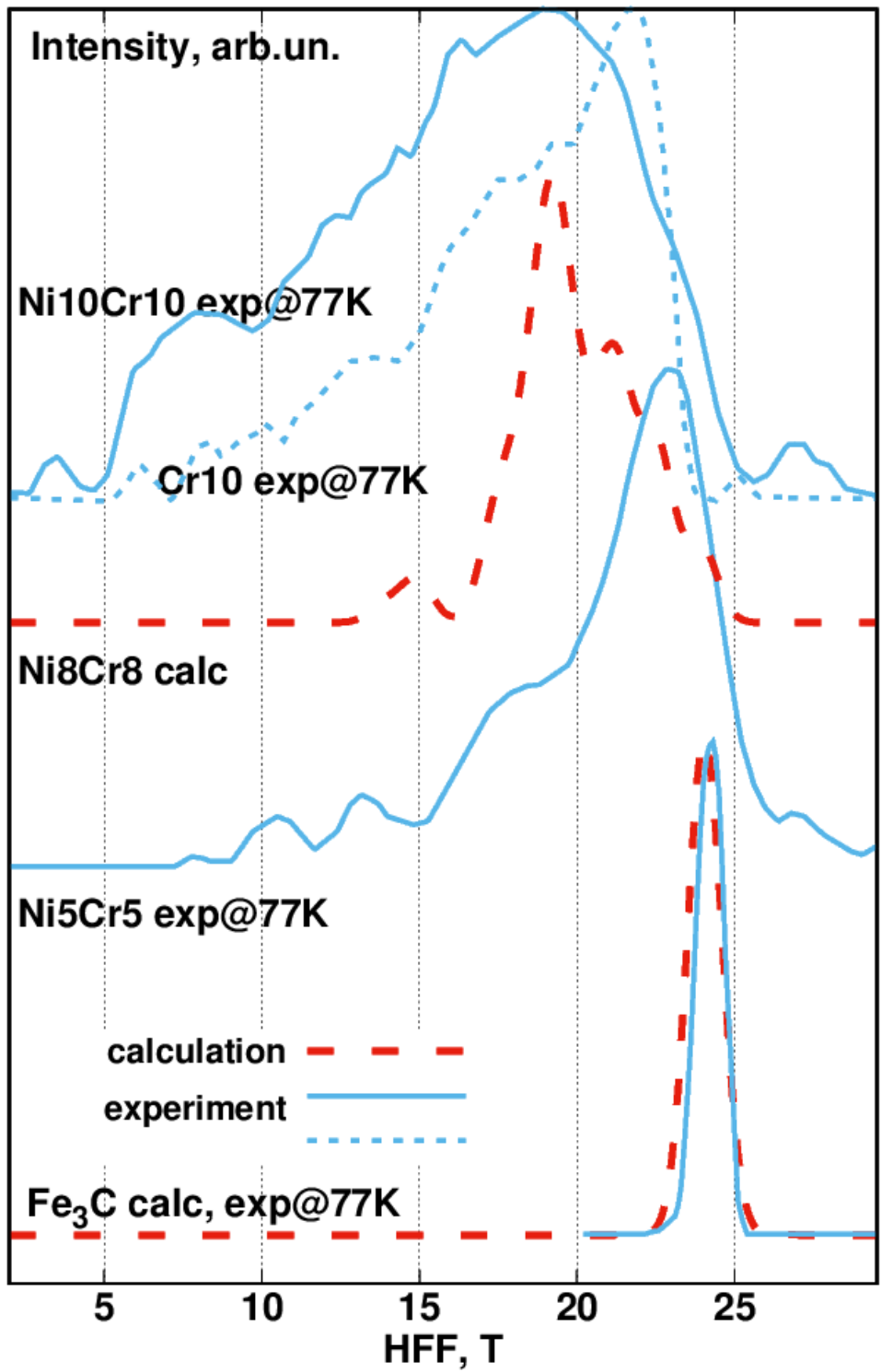

Fig. 12. Experimental [3,6-8] and simulated P(H) for Ni+Cr-doped cementite compared with pure $Fe_3C$, experimental P(H) of Cr10 is also shown (notations as in Fig. 10).

Both experimental and calculated *P*(*H*) shift toward lower HFF and broaden with increasing Ni and Cr content. For low concentrations (Ni5Cr5), the experimental *P*(*H*) closely matches the calculated Ni8Cr8, but at higher concentrations (Ni10Cr10), it becomes significantly broader. The experimental Cr10 *P*(*H*) (dashed blue line), along with Ni15 and Ni17 (Fig. 11), cannot account for this broadening, suggesting an unclear structural transformation. Magnetic measurements in this system also revealed anomalies [8], which its authors linked to the formation of two phases with differing Cr content. This phase separation aligns with Cr carbide's known segregation tendency [1], consistent with chromium's role as a carbide-forming element.

## 4. Conclusions

This study investigates cementite alloying with Ni and Cr using DFT calculations, focusing on their differing carbide-forming abilities and magnetic behaviors. We analyze structural parameters and hyperfine fields in $(Fe-Ni-Cr)_3C$ systems, highlighting the effects of Ni and Cr alloying on hyperfine magnetic fields. The calculated results are validated against previous experimental data from XRD and Mössbauer spectroscopy studies of mechanically alloyed samples [2-8].
1. Comparison of calculated and experimental structural parameters reveals that high-energy mechanical alloying enforces random Ni distribution across general and special positions, with no subsequent redistribution occurring after 500°C heat treatment (Figs. 3-4). For the Cr system, available data remain insufficient to determine preferential Cr site occupation (Figs. 2, 4).
2. The Fe hyperfine field and magnetic moment show no strict correlation (Fig. 6), as the valence electron contribution is highly environment-dependent, while the core contribution to HFF is nearly proportional to the Fe magnetic moment. The observed HFF dispersion reaches 7 T for identical moments. Conversely, for a given HFF value, the magnetic moment varies within ±0.2 $\mu_B$.
3. The commonly assumed relationship between hyperfine field and nearest-neighbor impurity number does not hold for the $(Fe-Ni-Cr)_3C$ system (Fig. 7). Significant HFF variations occur even with identical numbers of nearby impurities, making standard discrete-model analysis of Mössbauer spectra incorrect.
For Cr-alloyed cementite at low concentrations, spectral fitting remains possible using average HFF values and their variation: $<H^0>$=23.5T ±1T, $<H^1>$=20.5T ±2T, $<H^2>$=18T ±3T (note partial overlap between $P(H^1)$ and $P(H^1)$).
In contrast, Ni-doped systems show strong overlap even at low concentrations: $<H^0>$=23.5T ±1.5T, $<H^1>$=22.5T ±2T, $<H^2>$=22T ±2T. This makes $H^0$ and $H^1$ particularly difficult to resolve.
4. The continuous analysis of Mössbauer spectra often assumes a linear HFF-IS correlation. However, Fig. 8 demonstrates this approximation's significant inaccuracy, and its application for inverse problem simplification may introduce unpredictable distortions.
5. Comparison of calculated and experimental *P*(*H*) distributions reveals that Ni alloying in cementite reduces the average HFF and broadens the distribution (Fig. 11). Cr alloying produces a similar but more pronounced effect (Fig. 10). At higher concentrations (10% Cr), the experimental *P*(*H*) broadening slightly exceeds theoretical predictions. Additional alloying with 10% Ni causes even greater broadening than expected from a simple additive model (Fig. 12). This suggests that Ni may promote phase separation into Cr-rich and Cr-poor regions, consistent with earlier hypotheses [8].

**Data availability**

Data will be made available on request.

**Acknowledgements**

The calculations were partly performed using the Uran supercomputer of the Institute of Mathematics and Mechanics, Ural Branch, Russian Academy of Sciences.
*Funding: This work was supported by the* Ministry of Science and Higher Education of the Russian Federation [no. 24021900079-9].

**Conflicts of Interest**

The author declares that there is no conflict of interests or personal relationships that could have appeared to influence the work reported in this article.

**Author contributions statement**

The author L. V. D. did everything in the paper herself.